\input phyzzx
\centerline{\bf NONPERTURBATIVE INSTABILITY OF BLACK HOLES}
\vskip .2cm
\centerline{\bf IN QUANTUM GRAVITY}
\vskip .5in
\centerline{Pawel O. Mazur}\foot{E-mail address:
mazur@psc.psc.sc.edu}
\vskip .2in
\centerline{Institute for Fundamental Theory, Department of Physics,}
\centerline{University of Florida, Gainesville, FL
32611}\foot{Present address: Department of Physics and
Astronomy, University of South Carolina, Columbia, SC 29208}
\vskip .2cm
\centerline{September 1988}
\vskip .1cm
\centerline{Published in Mod. Phys. Lett.{\bf A4}, 1497 (1989)}
\vskip 2cm
\centerline{\bf Abstract}
\vskip .2in
\par
The final stage of a black hole evaporation due to the Hawking effect is
studied. One finds that, including the effects of quantum gravity, a
black hole does not evaporate completely losing its energy steadily
to a flux of created particles, but rather decays via a change
in topology into an asymptotically flat space and an
object which is a closed Friedmann Universe.
This process is a genuine nonperturbative effect of quantum gravity and
becomes the dominant ``channel'' of a black hole decay for black holes
with masses slightly larger than the Planck mass $M_{p}=10^{19}$GeV.
We calculate the decay rate of a Schwarzschild black hole with the mass
$M$ and discuss other decay ``channels'' by topology change. An explicit
instanton mediating the decay is constructed by matching
the Schwarzschild
and the ``wormhole'' Friedmann instantons on the minimal sphere which is
a Euclidean section of the event horizon. We show, as an example,
that the decay process is mediated in the semiclassical approximation
by the gravitational-axionic instanton. However, we argue that the
phenomenon discussed in this Letter does not depend on the particular
instanton approximation and should be discussed in the framework of the
second quantization of interacting geometry suggested in Ref. 18.
It is argued that in the more general setting of
the Wheeler-De Witt equation
the wave functional describing a black hole
is not gaussian because of the
existence of an unstable mode.
\vskip 2.5cm
\par
Several years ago Hawking discovered that black holes are quantum
mechanically unstable [1]. In the strong gravitational field of a black
hole, the in- and out-vacua of different matter fields are different
even if the object is characterized
by the stationary gravitational field.
It is the presence of the event horizon, whose sections are
topologically $S^2$, that makes the difference.
The nontrivial topology
of black holes and their highly nontrivial
quantum mechanical properties
is probably the strongest argument in favor
of considering quantum
fluctuations in the topology of spacetime.
The general concept
of quantum fluctuations in the metric
(the perturbative sector)
and a possible topology change with the ``pinching off''
of separate Universes (or regions of spacetime) was first discussed
by Wheeler [2], and more recently by Hawking [4,6].
Zel'dovich [3,5] has discussed the role of
topology change and its implications for particle physics
and cosmology. He also suggested that small black holes with masses
close to the Planck mass $M_p$ would decay in one quantum jump and a
small closed world will be formed in such a process [5] together with an
asymptotically Minkowski spacetime. Zel'dovich
envisaged that such a process would necessarily violate the baryon and
lepton number conservation law. Hajicek [16] in series of penetrating
and very beautiful papers addressed
the issue of the quantum mechanics of
spin-zero
black holes. He and his collaborators
quantized the spherically symmetric
gravitational and matter fields in
the topological black hole sector.
Hajicek was able to solve the hamiltonian
contraints and derived the
reduced highly nonlinear but positive
hamiltonian for the solitonic
(extremal) black holes and matter fields.
Indeed, this is the only example
known to me where the serious attempt was made to quantize black
holes and matter fields beyond the semiclassical approximation.
\par
The physical implications of topology of black holes were
recently discussed in the context of string theory [7].
In particular the fact that the second homology of a black hole
manifold is nontrivial implies the existence of string world-sheet
instantons whose role is to destabilize a black hole.
The general picture which emerged from this work is that
very small black holes
as described by string theory undergo
a sort of the phase transition to
a different phase. When the Hawking temperature becomes large and
comparable to the Hagedorn temperature, the string partition function
at genera zero and one diverges, signaling instability. The effects of
string world-sheet instantons and vortices lead to instability at genus
zero. The effects of the Wess-Zumino-Novikov-Witten term
(the two-form $B$ field) on a black hole dynamics in string theory
were discussed and a critical value for the mass of
a black hole was found [7].
\par
Indeed, a very small black hole of mass $M$ comparable to the Planck
mass $M_p$ has a very high Hawking temperature $T={M_p}^2/8{\pi}M$.
The average thermal energy of the particles emitted by such black holes
is slightly below the Planck mass, and it is definitely favorable for
a black hole to ``disappear'' in a quantum fluctuation (as a discrete
``quantum jump''), with a possible change of topology [5].
Also the average number $N$ of particles
produced in the black hole decay
is of order $N\cong 8{\pi}M^2/{M_p}^2$.
Notice that this number is
proportional to the geometrical scattering
cross-section or the area of a
black hole horizon. Imagine now that
a black hole is formed as an intermediate
state in the collision of high energy
elementary particles, say baryons.
Such a state would decay rapidly producing a number of other particles.
We do not know the hamiltonian describing
such a process, because we do not
have yet the microscopic fundamental theory of all interactions.
On the other hand we know that the multi-particle production in the
high energy collision of hadrons can
be phenomenologically described by the
thermodynamical models. Such an approach to hadron collisions
proved to be moderately succesful even
if the average number of produced
particles was not very large.
One would expect that such an approximation
becomes more accurate as the
number of produced particles increases.
The metastable intermediate
state tends to behave thermodynamically as the
number of ``fragments'' increases,
which indicates that, even if the quantum
coherence is not lost in such a process,
the phase space becomes very large.
In the fundamental theory of all interactions
quantum black holes would
probably appear as such ``resonances'', or collective excitations,
which eventually would ``fragment'' into a number of particles. The fact
that the quantum mechanical decay of such a state can be described by
thermodynamics simply reflects the hierarchy problem in quantum gravity
(QG). Indeed, the mass scale of QG $M_p$ is much above the energy
scale of other fundamental interactions.
It seems that this property of QG
might be responsible for the large number
of particles produced in the
decay of a quantum black hole.
\par
The nonperturbative instability of black holes by topology
change will become the dominant channel of decay
for very small Planckian black holes. The usual Hawking effect decay
mode of particle creation probably becomes nondominant as the genuine
quantum gravity (QG) effects start to dominate. The standard picture of
final stages of the black hole evaporation does not seem to take into
account the quantum effects of gravity.
In particular, it seems unphysical
to suggest that a black hole evaporates
completely and that the resulting
spacetime will possess a ``naked singularity''. The final stage of a
black hole evaporation due to the Hawking effect is a unique laboratory
for any theory of quantum gravity. The Einstein-Hilbert action
may not be a very good approximation of QG at the Planck energy scale
$M_p$, but presumably one may consider it as the leading term in the
gradient expansion of the low energy effective action for energy scales
slightly below the Planck energy.
\par
This assumption validates the use of the Wheeler-De Witt (WDW)
equation [9] for the description of small ``quantum black holes''.
Consider the real time, i.e., Lorentzian, configuration of
the gravitational field, say a black hole.
In the quantum mechanics of the gravitational field [9-12],
one associates a wave function to this configuration.
It is known how to do this, at least semiclassically,
in a WKB approximation. One considers a Riemannian
three-manifold $\Sigma$ which is
an initial data hypersurface $\Sigma$ in classical
general relativity (GR). The initial data is a canonical pair
$(h_{ij},{\pi}_{ij})$, a ``point'' in the phase space where the
semiclassical wave function is localized.
This assumption is based on the analogy with quantum mechanics
of simpler systems. Usually one
chooses the polarization on the classical phase space such that
the WDW wave function is a functional of the three-metric $h_{ij}$ :
${\Psi}[h_{ij}]$.
\par
By the choice of the classical solution or, equivalently, by chosing
a specific initial data on $\Sigma$, one associates with it the
semiclassical wave function in the classically allowed region of a phase
space. Now a given classical solution is a critical point of the action.
By expanding the action around a classical solution one finds in the
functional integral approach the gaussian wave function of a ``ground
state'' [11,15,20]. The other way to do it is
to solve the WDW wave equation and find the $O(h^0)$ quantum correction
to the WKB wave function.
\par
A ``ground state'' wave function may cease to be gaussian in some
directions in the configuration space [15,20]. This behavior signals
instability. Such a ``ground state'' is a ``false vacuum''. It tends to
decay, i.e., there exists an ``overlap''
between the ``false vacuum'' and
another ``vacuum''.
The wave function tunnels to a classically forbidden
region. Such a tunnelling is most conveniently described
in the path integral approach where one can calculate the transition
amplitude in a WKB approximation.
This leads automatically to an instanton
mediating such a transition.
In this Letter we are interested in studying
the semiclassical nonperturbative instability of ``small'' black holes.
\par
In general relativity a black hole is stable. The topology of $\Sigma$
does not change in classical general relativity because it would lead
to causality violation. Therefore, the system under consideration
tunnels to another classically allowed region of the phase space if a
semiclassical instability is really present. In general relativity
(GR) the tunnelling can be associated with topology change [8,9,20].
This can be seen heuristically in the Feynman sum over histories
approach where one sums over all possible ``histories''
with the amplitude ${\rm exp}(iS[g])$ as the weight factor,
where $S[g]$ is the classical action. The ``history'' corresponding
to topology change does not correspond
to a solution of the classical Einstein equation (here we define the
solution as the Lorentzian geometry with the smooth light cone structure,
at least outside event horizons). As a result the action
will not have an extremum, and in the sum over histories the total
amplitude will be exponentially small because of the effects of
destructive interference. This leads to the Euclidean
continuation of the action in the semiclassical approximation
and to the Euclidean QG prescription.
For our purposes in this Letter we simply assume that the standard
Euclidean arguments are applicable (this may not be true beyond the tree
level).
\par
Studying the semiclassical instability of black holes due to topology
change in QG, one must find an instanton which mediates this
instability. The classical ``ground state'' of a black hole is uniquely
described by the Schwarzschild solution, which has the
topology $R^2\times S^2$ .
The presence of an instability can be established most easily by
analytically continuing this solution to a Euclidean space signature, so
that the metric is
$$ds^2=(1-2GMr^{-1})d{\tau}^2+(1-2GMr^{-1})^{-1}+
r^2d{\Omega}^2, \eqno(1)$$
where $M$ is a black hole mass and $G={M_p}^{-2}$ is the Newton constant
given in terms of the Planck mass $M_p$. The Schwarzschild radius is
$R=2M{M_p}^{-2}$. The constant time
section of this geometry has topology $R\times S^2$. This is the
famous Einstein-Rosen bridge or the three-dimensional wormhole connecting
two asymtotically flat regions.
If one restricted the range of the radial
coordinate to $R<r<\infty$, this three-geometry would be an
incomplete manifold. However, one can see that the origin of this
incompleteness is the presence of a ``hole'' at $r=R$. This ``hole''
is a minimal $S^2$ on the manifold and one can double the manifold by
gluing in the exact copy of it at the minimal $S^2$. In this way one
obtains the Einstein-Rosen bridge. This procedure corresponds to the
maximal analytic extension of the Schwarzschild solution and is not
dictated by the laws of physics. Evidently the realistic black hole
formed in a gravitational collapse is asymmetric in this respect.
Quantum mechanically a black hole radiates particles, out of
the vacuum, which escape to the asymptotic region. Now this positive
energy flux of outgoing particles is always accompanied by a
negative energy flux crossing an event horizon. The situation is time
asymmetric.
Therefore, physically the initial constant time section is only
half of the Einstein-Rosen bridge. One can find a coordinate system on
half of the wormhole such that the metric is conformal to the metric
on half of the three-sphere $S^3$. However, the conformal factor will
be singular at one point,
this corresponding to the fact that the topology
of the brigde is different than a three-dimensional disk $D^3$.
The singular point on this disk is the ``infinity'' on the bridge.
One sees that by compactifying the bridge by adding a point at infinity,
or equivalently by conformally rescaling the metric, we obtain a
three-manifold which is topologically $S^3$,
with the boundary $S^2$ being
a minimal two-sphere.
\par
How do we search for a semiclassical instability of some
given configuration?
The proper tool we need to address the question of topology change is
cobordism theory. A general discussion of this approach was presented
in Ref.(9) . Let $\Sigma$ be an initial
three-geometry and $\Sigma'$ be a final three-geometry. The question of
semiclassical instability of $\Sigma$ can be formulated now as the
problem of the existence of a smooth Riemannian manifold $\cal M$
interpolating between $\Sigma$ and $\Sigma'$. Two oriented manifolds are
called cobordant if their disjoint union bounds a smooth manifold,
$\partial{\cal M}=\Sigma\cup\Sigma'$. The basic result of cobordism
theory which is useful here is that two manifolds are cobordant if
their Pontryagin and Stiefel-Whitney numbers are equal. This implies in
particular that all closed oriented three-manifolds are cobordant.
This also means that $S^3$ can ``decay'' into any closed arbitrarily
complicated oriented three-manifold. If a manifold is compact
and has a boundary then one can modify the present argument
and consider cobordisms with fixed boundaries.
\par
Now a three-sphere with a boundary $S^2$ is cobordant to any oriented
three-manifold with $S^2$ boundary. This means that, at least in
principle, a black hole can
decay into any topologically nontrivial configuration. Consider, as in
Figure 1, the complete
three-manifold obtained by ``filling in'' a hole in $R^3$, i.e., by
gluing in a disk $D^3$ (or $S^3$ with a boundary $S^2$) to half of the
Einstein-Rosen bridge: $R\times S^2 + D^3 = R^3$.
Adding a sphere $S^3$ to
a disk $D^3$ ($D^3\cup_g S^3\equiv D^3$)
does not change topology of a disk:
$R\times S^3 + D^3\cup_g S^3= R^3 + S^3$,
but rather corresponds to
a process of producing a disjoint union of $R^3$ and $S^3$.
This ``cupping'' or ``plumbing'' operation produces
a complete manifold $R^3$ (or $R^3 + S^3$).
The procedure described above
is performed on the constant time initial data three-geometry.
One would like to find an instanton solution
corresponding to this decay mode.
The simplest possible way to obtain such
an instanton is to match a Euclidean
black hole solution to a Friedmann
universe on the constant time hypersurface.
However, there are no vacuum solutions corresponding to this
``match''.
Consider a four manifold ${\cal M}_{F}$ of topology $S^3\times
R$ with a minimal $S^3$. Cut a wormhole ${\cal M}_{F}$ in half
on the minimal $S^3$. Next cut this minimal $S^3$, a constant ``time''
slice of a wormhole geometry, along an equator $S^2$ obtaining thereby a
disk $D^3$. A disk $D^3$ is topologically the same as a disk $D^3$ and a
sphere $S^3$ glued together: $D^3\equiv D^3\cup_g S^3$.
Now glue the constant time slice of
the Euclidean Schwarzschild spacetime
(ES) along the ``horizon'' $S^2$ to
an equator of the minimal $S^3$ of
the half-wormhole.
This operation defines the manifold of an instanton
which mediates the decay of a black hole
to a closed Friedmann universe and
an asymptotically flat spacetime without a hole.
The hybrid four-manifold
obtained this way is the ES on the one side of
the hypersurface of a constant
time and the Friedmann universe on the other side.
\par
Now we have to find a solution to the Euclidean equations of motion
corresponding to the process of ``pinching off'' a small closed
universe. Indeed, one can find such an instanton solution for
the Kalb-Ramond two-form $B$ coupled to gravity.
There are other ways to obtain such an instanton, but from
now on we focus our attention on this simple model. We simply observe
that the Bekenstein result [14] which says that a static black
hole does not have scalar (spin $0$) hair can be easily extended
to the Euclidean instanton case. The instanton solution in question is
simply the ES on one part of an instanton manifold
and a four-dimensional wormhole on the other part of a complete
instanton manifold. One simply requires smooth matching
conditions on the metric and the second fundamental
form of a matching surface. Indeed, one can show that
such a smooth matching of the ES and the Friedmann universe does exist.
In any case one need require only that
the action of an instanton is finite.
\par
The Euclidean Schwarzschild solution is known to possess the amazing
property of having one normalizable negative mode in the ``graviton''
sector. What does this mean physically?
Gross et al. [13] interpreted this fact as corresponding to a
semiclassical nonperturbative instability of flat Minkowski spacetime
in a thermal bath due to the nucleation of black holes
(in contrast to the perturbative Jeans instability; see also Ref. (17)
for the another point of view). Mathematically,
this means that the ES instanton is only
a local extremum of the Euclidean
Einstein-Hilbert action. The possibility of the
``phase transition'' of a black hole
in the ``box'' (in the ``heat bath'')
with a possible topology change was suggested in the
different context in Ref. (17).
\par
The four-dimensional wormhole (the Einstein-Rosen bridge)
simply has the Euclidean Friedmann
metric which is the solution of
the coupled axion and Einstein equations.
One can show that this solution is absolutely stable
(in analogy to the real time Friedmann universe which is stable).
This is implied by the positive action theorem.
In the linear approximation,
one finds no negative modes in the spectrum
of deformations of the wormhole.
On the other hand the half-wormhole does have negative modes in its
spectrum of deformations.
Observe that the time-symmetric initial value data corresponding to
the 5-dimensional static Tangherlini
black hole also describes a wormhole.
However, this wormhole cannot be smoothly matched to the ES instanton.
One of the interesting properties of this wormhole, which deserves
a separate note, is the fact that the asymptotically Euclidean (AE)
gravitational instanton
with the topology $R^4\cup_g{\cal M}$ and zero Ricci scalar $R$ has the
Euclidean action greater or equal to the action of the Tangherlini
wormhole. Indeed, one can show that the action of the AE gravitational
instanton whose manifold is $R^4$ outside
a bounding $S^3$ is not less than
that of the Tangherlini instanton,
with the equality achieved only when the
metric is that of the Tangherlini wormhole
(Mazur 1987, unpublished).
\par
The instanton solution described above mediates the decay
of a black hole into a small closed universe ``pinching off'' from our
``large'' Universe containing a black hole.
This instanton must correspond
to a local extremum of the Euclidean action.
This means that there exist negative modes in the spectrum
of fluctuations around this instanton.
Indeed, a wormhole cut in half and
the Euclidean Schwarzschild do have negative modes. Therefore, the
``matched'' instanton solution corresponding to topology change
$S^2\times R + D^3$ to $S^3 + R^3$, and from $R^2\times S^2$
to $R\times S^3 + R^4$, has
negative modes in the ``graviton'' sector. In fact, there is only one
normalizable negative mode around this instanton [15].
\par
It is sufficient to present here the asymptotic form
(as $\tau\rightarrow -\infty$) of the metric on the
wormhole cut in half
$$ds^2=d{\tau}^2+{a^2}({\tau})d{\Omega_3}^2 ,$$
$$a^2({\tau})\cong{\tau}^2(1+2R^4/{3{\tau}^4}) . \eqno(2)$$
Indeed, one shows that
the exact solution to the coupled Einstein
and the axion $B$-field equations
can be given in the parametric form
$$a({\eta})=R(cosh2{\eta})^{1/2}$$
$${\tau}=\int{d{\eta}a({\eta})} , \eqno(3)$$
and $dB=g(\tau){\epsilon}$, where
${\epsilon}$ is the volume 3-form on $S^3$.
>From the axion equations of
motion: $d*H=0$, $H=dB$, one finds
$$g=q/2{{\pi}^2}{f^2}a^3({\tau}) , \eqno(4)$$
where $q$ is an integer global axion charge.
Here $d{\Omega_3}^2$ is the metric on a unit
round sphere $S^3$ and $R$ is the
radius of the minimal $S^3$ which equals the
Schwarzschild radius. Indeed,
this radius $R$ is determined by the condition
that the Euclidean
Friedmann solution matches the Euclidean
Schwarzschild solution. $R$
depends on the axion coupling $f$ and the global charge $q$.
We will not need the details
of this relation here (this will be discussed elsewhere [15]).
Anyway it is a rather trivial excercise.
One finds the Euclidean action of the wormhole:
$I_W=0$, whereas the action
of the ES instanton is: $I_{S}=\pi{M_p}^2R^2$.
The total action of the instanton is
$$I=I_{S}+I_{W}={\pi{M_{p}}^2R^2} , \eqno(5)$$
because the action of the wormhole is zero.
In terms of the black hole mass, the action is
$I={4\pi{M^2}}/{{M_p}^2}$, where we have used $R=2M{M_p}^{-2}$.
\par
Now we can estimate a decay rate of a black hole with a mass $M$.
Indeed, in the dilute instanton gas approximation
one finds the decay rate
per unit (Planck) volume [19]
$$\Gamma=A{\rm exp}(-I) , \eqno(6)$$
where $A$ is the imaginary part of the functional determinant
of the small fluctuation operator around an instanton.
For the case under consideration,
this is a determinant of the Lichnerowicz laplacian
${\Delta_{L}h_{ab}}=-{\nabla}^2h_{ab}+2{{R_a}^{cd}}_bh_{cd}-
2R_{c(a}h^c_{b)}$ (in the more careful analysis
one would also include the
contribution from the axion field fluctuations).
For our purposes we need only the order of
magnitude estimate of $A$, $A\cong O(1){M_p}^5{M}^{-1}$.
The decay rate of a black hole thus equals
$$\Gamma\cong O(1){M_{p}}^5{M}^{-1}{\rm exp}(-{4\pi{M^2}}
/{{M_{p}}^2}) . \eqno(7)$$
For a large black hole the decay rate due to this nonperturbative
instability is extremely small and the decay time is much larger
than the age of the Universe. However, for the mass of a small
black hole, $M={M_p}$ the decay rate per unit spacetime volume
is only of order $10^{-6}$. The spontaneous decay rate might indeed be
small but the decay stimulated by the environment might be much higher.
Indeed, this is what we may expect to happen
in the very early post-Planckian
Universe. The mechanism described in this Letter
might explain the absence of primordial black holes.
Indeed, it
seems that unlike the monopole problem the primordial
black hole problem is
easier to resolve because black holes are unstable,
while monopoles are
stable.
\par
It should be noticed that the decay rate of
a black hole due to the nonperturbative instability
is proportional to
$exp(-S_{bh})$, where $S_{bh}$ is the Bekenstein-Hawking entropy.
This seems to indicate that the Bekenstein-Hawking
entropy is the measure
of the ability of a black hole to disappear from our Universe in the
one quantum jump, with the simultaneous production of particles carrying
away its total energy. Indeed, the naive estimate of a number of
microstates which correspond to a given black hole of the mass $M$ is
$N\cong{\Gamma}^{-1}$. The microcanical entropy for such a state is

$$S_{bh}={lnN}\cong{-ln{\Gamma}}=4{\pi}{M^2}{M_{p}}^{-2} . \eqno(8)$$

Zel'dovich has shown long time ago (in 1962, see Ref.(5)),
that any number
of baryons with any entropy can be brought into a configuration with an
arbitrarily small rest mass.
This is because the gravitational mass defect
can almost completely compensate the total energy of the matter
configuration. This implies that any matterial object is in fact
metastable, or unstable against the decay mode with the transition to
the more condensed state with an excess of energy being radiated away to
infinity in the form of radiation.
The apparent stability of matter is just
a reflection of the fact that there
exists an enormous energy barier
through which the system must tunnel to the superdense state.
But the same principle also seems to apply to black holes.
The Zel'dovich
mechanism through which a number of baryons disappears
from our Universe
into a Closed World with the simultaneuos production
of particles with a
zero net baryon number and the same total energy can proceed in two
different channels. One of them is a formation of
a wormhole through which
baryons disappear to another Universe,
and the other is through the
formation of a black hole which eventually
decays via the mechanism
described in this Letter.
Both processes lead to the baryon number
nonconservation.
\par
In order to demonstrate the idea
of the nonperturbative instability of
black holes in QG it was sufficient to find only one decay mode
of a black hole. The decay mechanism which becomes dominant
at the Planck scale is due to topology change.
There are other decay modes for
which we have not constructed instantons.
It is the total decay rate of
all these `` channels'' which will give the correct decay time for a
small black hole. Does this mean that one has to know all the decay
channels to different topologies in order to calculate the total
width of a quantum black hole state?
One would expect that there will exist dominant channels for
the decay such as the one we have demonstrated above. Evidently it is
not clear at present to what degree one can trust the semiclassical
calculations like those presented above. One has to bear in mind that
the proper calculation of the decay rate $\Gamma$ should be done in the
hamiltonian formulation, i.e., using the WDW equation for the
minisuperspace incorporating the properly matched Friedmann
and Tolman metrics.
The Euclidean approach to the tunnelling process
described above (i.e., a black hole decay)
could only be justified after
the correct hamiltonian calculation was done
(see, e.g., Ref.(15)). This
approach will be described elsewhere [15].
In any case in this Letter we
have presented not only the qualitative
picture of the nonperturbative
quantum gravitational decay of black holes but also a quantitative
estimate of the decay rate for such a process.
A more detailed discussion of the result presented here will be
given in a forthcoming paper$^{15}$
where the Wheeler-De Witt equation for
``isolated'' gravitating objects
is derived and applied to the quantum
mechanics of black holes.
\par
I am grateful to Petr Hajicek who in Summer of 1984 explained to me
the results of his deep and penetrating studies
of the quantum gravitational collapse.
\par
I would also like to thank Henry Kandrup, Emil Mottola, V. P. Nair,
James York Jr., and Frank Wilczek for their interest in this work.
I am especially indebted to Mark Bowick,
who was always willing to listen and discuss with me when I was working
on black hole topological string solitons during the winter 1986/87.
\par
My work was supported by NSF grant to Syracuse University
Physics Department Relativity Group and only partially (1/2)
by the Institute for Fundamental Theory,
Department of Physics, University of Florida, Gainesville.
\endpage
\vfill
\doublespace
\vskip .2in
\centerline{REFERENCES}
\vskip .2in
\item{^1}      S. W. Hawking, Comm. Math. Phys. {\bf 43}, 199 (1975).
\item{  }      R. M. Wald, Comm. Math. Phys. {\bf 45}, 9 (1975).
\item{  }      J. B. Hartle and S. W. Hawking, Phys. Rev.
{\bf D13}, 2188 (1976).
\item{^2}      J. A. Wheeler, Ann. Phys. {\bf 2}, 604 (1957).
\item{^3}      Ya. B. Zel'dovich, in General Relativity,
An Einstein Centenary Survey, eds. S. W. Hawking and W. Israel
(Cambridge University Press, Cambridge 1979).
\item{^4}      S. W. Hawking, Phys. Lett. {\bf 195B}, 337 (1987).
Also Ref. (3).
\item{^5}      Ya. B. Zel'dovich, Usp. Fiz. Nauk. {\bf 123}, 487 (1977).
[Sov. Phys. Usp. {\bf 20}, 945 (1978)].
\item{^6}      S. W. Hawking, Phys. Rev. {\bf D37}, 904 (1988).
\item{^7}      P. O. Mazur, GRG {\bf 19}, 1173 (1988).
\item{  }      P. O. Mazur, ``The Hawking Effect in String Theory and the
Hagedorn Phase Transition'', talk given at Yale University, November 1987
(unpublished).
\item{^8}      F. A. Bais, C. Gomez and V. A. Rubakov, Nucl. Phys. {\bf
B282}, 531 (1987).
\item{^9}      P. O. Mazur, Nucl. Phys. {\bf B294}, 525 (1987).
\item{^{10}}   B. S. DeWitt, Phys. Rev. {\bf 160}, 1113 (1967).
\item{^{11}}   C. Teitelboim, Phys. Rev. {\bf D25}, 3159 (1982).
\item{^{12}}   J. B. Hartle and S. W. Hawking, Phys. Rev. {\bf D28}, 2960
(1983).
\item{     }   K. Kuchar, J. Math. Phys. {\bf 11}, 3322 (1970).
\item{     }   J. B. Hartle, Phys. Rev. {\bf D29}, 2730 (1984).
\item{^{13}}   D. J. Gross, M. J. Perry and L. G. Yaffe,
Phys. Rev. {\bf D25}, 330 (1982).
\item{^{14}}   J. D. Bekenstein, Phys. Rev. {\bf D5}, 1239 (1972).
\item{^{15}}   P. O. Mazur, in preparation. This paper contains the
Schr\"odinger
\item{     }   equation for black holes and ``geons''. This was also the
\item{     }   subject of my January 1989 Chapel Hill, UNC, Colloquium
\item{     }   and seminar. I have described this work to many people
\item{     }   including C. Teitelboim, L. Smolin and others.
\item{     }   P. O. Mazur, talk given at the 14th Texas Symposium on the
Relativistic Astrophysics, December 15, 1988.
\item{^{16}}   P. Hajicek, Phys. Rev. {\bf D30}, 1185 (1984); see also the
references to this author earlier work.
\item{     }   A. S. Goncharov and A. D. Linde, Fiz. Elem. Chast. Atom.
Yad., {\bf 17}, 837 (1986)[Sov. J. Part. Nucl. {\bf 17}, 369 (1986)].
\item{^{17}}   J. W. York, Jr., Phys. Rev. {\bf D33}, 2092 (1986).
\item{     }   B. F. Whiting and J. W. York, Jr., Phys. Rev. Lett. {\bf
               61}, 1336 (1988).
\item{^{18}}   P. O. Mazur and V. P. Nair, Gen. Rel. Gravitation,
in press (1989), ``An Interacting Geometry Model and Induced Gravity'',
Gravity Research Foundation Essay, March 1988. {\it This essay describes
the first physical application of the topological QFT, which is the
topology changing amplitudes in quantum gravity}. Published in
{\bf GRG21}, 651-658 (1989).
\item{^{19}}   S. Coleman, Phys. Rev. {\bf D15}, 2929 (1977).
\item{^{20}}   P. O. Mazur and E. Mottola, ``The Gravitational Measure,
Solution of the Conformal Factor Problem and Stability of the Ground State
of Quantum Gravity'', Los Alamos and University of Florida preprints,
submitted.
\item{     }   Published in Nucl. Phys. {\bf B341}, 187-212 (1990) with
the changed title:
\item{     }   ``The Gravitational Measure, Conformal Factor Problem and
Stability of the Ground State of Quantum Gravity''.
\item{     }   H. E. Kandrup and P. O. Mazur, ``Particle Creation and
Topology Change in Quantum Cosmology'', preprint UFT-88-10, submitted to
Nucl. Phys. {\bf B} but unpublished.
\vskip 1cm
\centerline{\it Selected References With Comments Added on September 29,
1997}
\vskip .5cm
\item{   1.}   H. E. Kandrup and P. O. Mazur, ``A Topological Hawking
Effect'', Mod. Phys. Lett. {\bf A4}, 1189-1196, (1989).
\item{     }   A very much shortened version of the last position in
Ref. 20 of the present paper. It discusses the effect of particle
creation on tunneling geometries (gravitational instantons) in the
context of cosmology and black holes.
\item{   2.}   P. Kraus, ``Hawking Radiation From Black Holes Formed
During Quantum Tunneling.'', Nucl. Phys. {\bf B425}, 615-633 (1994),
gr-qc/9403048.
\item{     }   A very nice paper which discusses tunneling of black
holes and particle creation on instantons.
\item{   3.}   P. Kraus and F. Wilczek, ``Selfinteraction Correction to
Black Hole Radiance'', Nucl. Phys. {\bf B433}, 403-420 (1995),
gr-qc/9408003.
\item{   4.}   P. Kraus and F. Wilczek, ``Effect of Self-Interaction
on
Charged Black Hole Radiance'', Nucl. Phys. {\bf B437}, 231-242 (1995),
hep-th/9411219.
\item{     }   Two very nice papers (Refs. 3,4), where the back-reaction
effects are taken care of in the WKB approximation. Direct relation to
tunneling is also quite clear. However, this work is really important
because it gives the clear derivation of the nonlinear dispersion
relation for quanta emitted by a black hole in the s-wave. This work
also leads to the new way of calculating the very high-energy tail
of the black hole decay in the D-brane context (Ref. 6 below).
\item{   5.}   P. Kraus, ``Nonthermal Aspects of Black Hole Radiance'',
Princeton University Ph.D. 1995 Thesis, 101 pp., gr-qc/9508007.
\item{   6.}   E. Keski-Vakkuri and P. Kraus, ``Microcanonical D-Branes
and Back Reaction'', Nucl. Phys. {\bf B491}, 249-262 (1997),
hep-th/9610045.
\item{   6.}   S. Massar and R. Parentani, ``Gravitational Instanton for
Black Hole radiation'', Phys. Rev. Lett. {\bf 78}, 3810-3813 (1997),
gr-qc/9701015.
\item{     }   I have learnt from this paper, (Ref. 6), that the formula
for the decay rate of a black hole due to
the nonperturbative instability in the
instanton approximation which was first drived in 1988 (in my paper
presented above), $\Gamma\sim e^{-S_{bh}}=e^{-{1\over 4}A_{bh}}$, was
also derived much later in Refs. 5,6 quoted in their paper.
I am not aware of earlier references which would relate
the black hole entropy to the decay rate of a black hole
due to tunneling. The exhaustive list of references
to this subject can be found in the 1995 Ph.D. Thesis of P.
Kraus (Ref. 5 above).
\endpage
\end